\documentclass[journal]{IEEEtran}
\usepackage{cite}
\usepackage{amsmath}
\usepackage{algorithmic}
\usepackage{array}
\usepackage{caption}
\usepackage{subcaption}
\usepackage{booktabs,tabularx,graphicx}
\usepackage{makecell}
\usepackage{graphbox}

\usepackage{color}
\definecolor{Blue}{rgb}{0,0,1}
\definecolor{Green}{rgb}{0,1,0}
\definecolor{Orange}{rgb}{1,0.5,0}
\definecolor{amethyst}{rgb}{0.6, 0.4, 0.8}
\definecolor{darkmagenta}{rgb}{0.55, 0.0, 0.55}
\definecolor{alizarin}{rgb}{0.82, 0.1, 0.26}

\begin{document}

\title{Evaluation of Neural Networks defenses and attacks using NDCG and reciprocal rank metrics}

\author{Haya~Brama, 
        Lihi~Dery, 
        and~Tal~Grinshpoun
\thanks{H. Brama, L. Dery and T. Grinshpoun are with the Department of Industrial Engineering and Management and Ariel Cyber Innovation Center, Ariel University 40700, Israel
e-mail: hayahartuv@gmail.com}
}

\maketitle

\begin{abstract}
The problem of attacks on neural networks through input modification (i.e., adversarial examples) has attracted much attention recently.
Being relatively easy to generate and hard to detect, these attacks pose a security breach that many suggested defenses try to mitigate.
However, the evaluation of the effect of attacks and defenses commonly relies on traditional classification metrics, without adequate adaptation to adversarial scenarios.
Most of these metrics are accuracy-based, and therefore may have a limited scope and low distinctive power.
Other metrics do not consider the unique characteristics of neural networks functionality, or measure the effect of the attacks indirectly (e.g., through the complexity of their generation).
In this paper, we present two metrics which are specifically designed to measure the effect of attacks, or the recovery effect of defenses, on the output of neural networks in multiclass classification tasks.
Inspired by the normalized discounted cumulative gain and the reciprocal rank metrics used in information retrieval literature, we treat the neural network predictions as ranked lists of results. 
Using additional information about the probability of the rank enabled us to define novel metrics that are suited to the task at hand.
We evaluate our metrics using various attacks and defenses on a pretrained VGG19 model and the ImageNet dataset.
Compared to the common classification metrics, our proposed metrics demonstrate superior informativeness and distinctiveness. 
\end{abstract}


\section{Introduction}
\label{sec:intro}
\IEEEPARstart{A}{s} research fields develop rapidly, evaluation metrics for these fields require adequate adaptation. 
Neural networks (NNs), especially in the form of deep learning, have achieved tremendous progress in pattern recognition. However, they remain vulnerable to attacks of adversarial examples (AEs). In such attacks, the attacker modifies the input, while carefully constraining the maximal size of modifications (perturbations) or the number of modified (perturbed) pixels in the original input image, so that the adversarial changes will be indistinguishable for humans. Due to these careful changes, the NN might incorrectly classify the AE~\cite{szegedy2013intriguing, nguyen2015deep, akhtar2018threat}. In this paper we introduce a new evaluation metric for the performance of artificial neural networks in the context of adversarial attacks. 

Attacks on NNs may be  effective even under strict constraints, such as a very narrow range of allowed perturbation size. Moreover, AEs can be transferred across different models. This opens the possibility of black-box attacks, where the attacker can create an effective AE even without access to the attacked NN's parameters~\cite{papernot2016distillation}. AEs may be robust to physical transformations~\cite{athalye2018synthesizing} and therefore pose a security breach not only for cyber-world applications~\cite{rosenberg2021adversarial} but also for real-world systems, such as computer vision of autonomous vehicles~\cite{evtimov2017robust}. Although various defense techniques have been suggested, none of them can ensure complete effective defense in all settings~\cite{goodfellow2014explaining, xu2018feature,ilyas2019adversarial}. 

This emerging research field requires adequate evaluation, in order to measure the effect of attacks and defenses on the functionality of the NN and to compare between different methods. 
Data classification literature employs several evaluation metrics in the pursuit of the most suitable or optimized classifier \cite{hossin2015review}. 
The evaluation metrics serve as a measurement tool for the generalization ability of trained models. 
Furthermore, evaluation metrics play a role during classifier training,
since they can guide a stochastic or heuristic search and help discriminate the optimal solution from the large space of solutions. 
The majority of those measures are based on some variation of the accuracy or the error-rate, as will be expanded below. Their simplicity makes them very intuitive and easy to use, but this high applicability comes with a price of rather limited informativeness and discriminative power~\cite{mackay2003information,wallach2006evaluation,huang2007constructing}. 
When the possibility of attacks and defenses was introduced in the literature, accuracy-based evaluation metrics were employed by default. 
However, as their original objective was to improve and assess the NNs' training in a benign setting, these measures are not necessarily the most appropriate choice for assessment in adversarial settings. 
As stated by Arp et al.~\cite{arp2020and}, many machine learning performance measures are not suitable in the context of security, since they may provide insufficient estimations or obscure experimental results. 
In addition, as various security-related problems deal with more than two classes, multi-class metrics are required and introduce further subtle pitfalls. 
Building on the observations of Arp et al., we argue that in many contexts current metrics do not always reflect the diversity of the adversarial scenarios, and fail to capture many of the strengths and weaknesses of attacks and defenses. 

\begin{figure}
     \centering
     \begin{subfigure}[t]{0.4\textwidth}
         \centering
         \includegraphics[width=\textwidth]{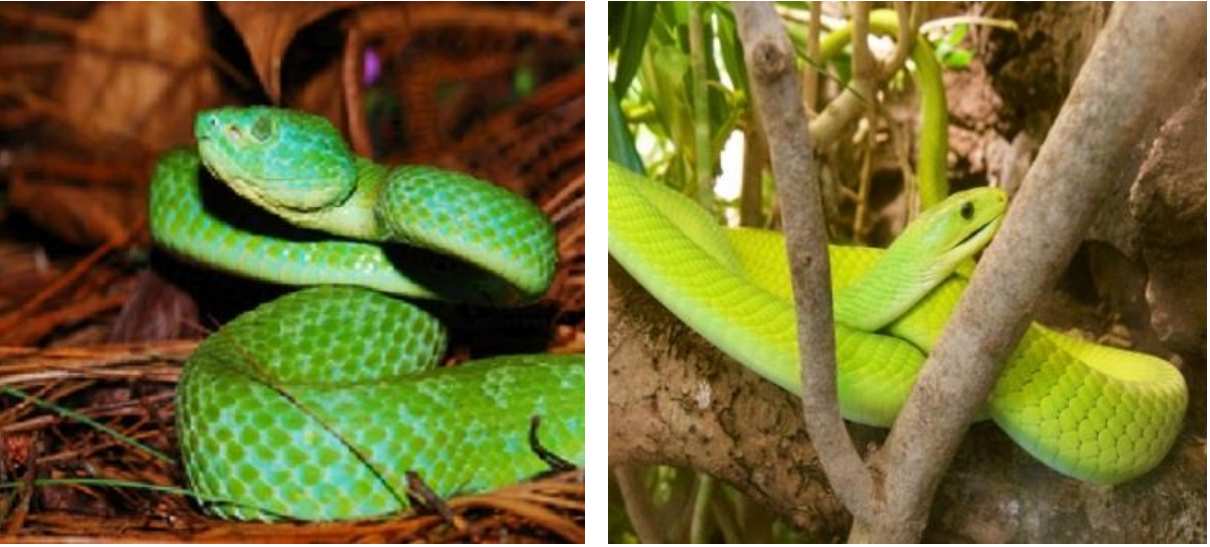}
         \caption{True label: green snake; Classified as: green mamba}
         \label{fig:minor misclassification}
     \end{subfigure}
     \hfill
     \begin{subfigure}[t]{0.4\textwidth}
         \centering
         \includegraphics[width=\textwidth]{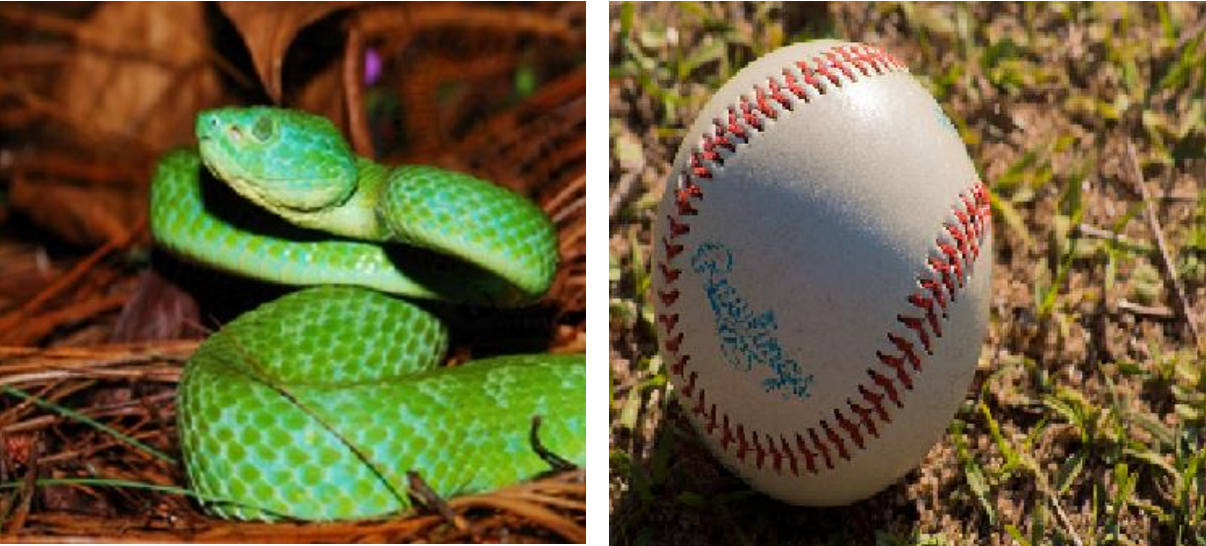}
         \caption{True label: green snake; Classified as: baseball}
         \label{fig:major misclassification}
     \end{subfigure}
     \hfill
     \begin{subfigure}[t]{0.4\textwidth}
         \centering
         \includegraphics[width=\textwidth]{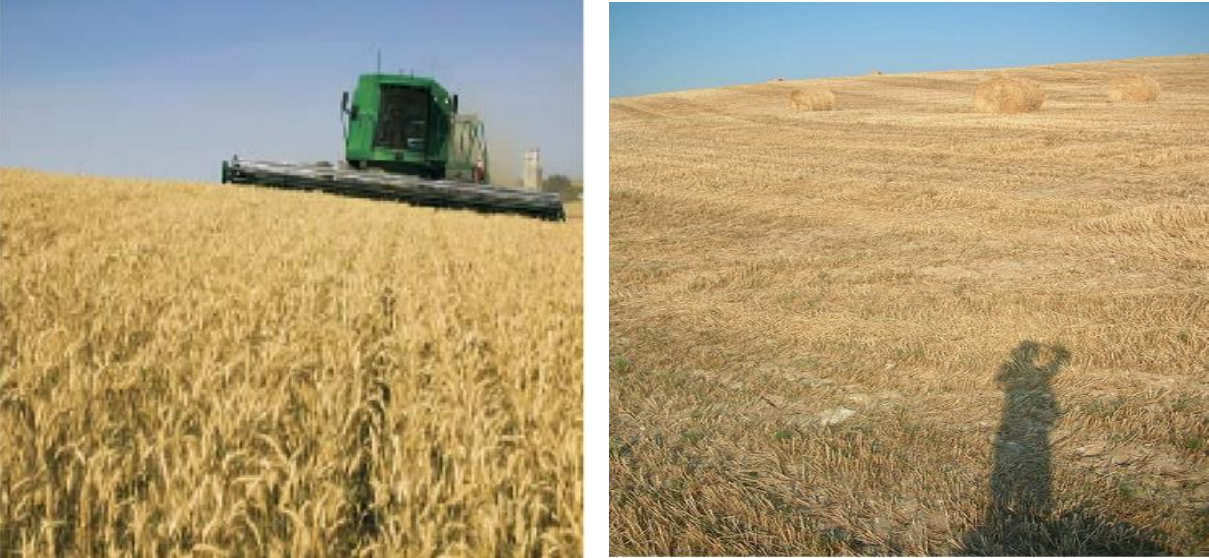}
         \caption{True label: harvester; Classified as: hay}
         \label{fig:wrong praioritization}
     \end{subfigure}
        \caption{Representative examples and their top-1 predictions that demonstrate three possible types of deceiving. (a) Minor misclassification: adversarial (left) and benign (right) images classified as `green mamba'. The original seed of the AE is classified as `green snake', and `green mamba' is the second highest prediction. (b) Major misclassification: adversarial (left) and benign (right) images classified as `baseball'. The original seed of the AE has `baseball' prediction in the 117th place. (c) Wrong prioritization: AE classified as `hay', where the image does present hay. The original seed is classified as `harvester', and `hay' is the second highest prediction. Images were taken from ImageNet and evaluated using VGG19.}
        \label{fig:deceiving types}
\end{figure}

To understand the limitations of current metrics, we first need to consider how they handle the NN's output.
The NN outputs a vector, whose length reflects the number of output labels. 
Each entry of that vector is proportional to the activity level of a different output neuron, which stands for a single category. 
The corresponding category of the most activated output neuron is selected as the NN's classification decision. 
The ratio between the highest activation and the overall activation of all the neurons is referred to as the confidence of the NN in its decision. 
Accuracy is usually measured as top-1 accuracy, which is the prevalence of the true label of the input matching the first-ranked predicted category.
Alternatively, accuracy can be measured more flexibly as the prevalence of the true label among the top-$k$ categories in the output vector, a measure known as top-$k$ accuracy.
Top-1 accuracy answers the limited question of \textit{``did the neural network output the exact correct label"?}
For example, according to top-1 accuracy, misclassifying the image `green snake' as `green mamba' is as wrong as misclassifying a 'green snake' image as 'baseball', see Figure~\ref{fig:minor misclassification}-\ref{fig:major misclassification}.
Even the top-$k$ accuracy measure does not always capture the whole classification picture, since it informs whether the correct label appeared in those top-$k$ predictions, but does not indicate where it is located within those predictions, or how far it is from their semantic neighborhood. 
There are many applications where this distinction is important, for example when the goal is to  visually identify weapons in an image for security reasons, and an attacker masks weapons as animals~\cite{athalye2018synthesizing}; in such a case, it is less crucial if a defense strategy turns the adversarial `animal' into one kind of weapon or another, as long as the object is now identified as a threat. 

Top-1 accuracy may cause another kind of misinterpretation. Consider a scenario in which an attack causes the NN to merely switch between the top-$k$ predictions. 
When an image consists of several objects, the resultant classification decision is not necessarily wrong per se.
For example, classifying the image labelled as `harvester' in the ImageNet dataset as `hay' (Figure~\ref{fig:wrong praioritization}) provides true information regarding to content of the image. 
To summarize, the current top-$k$ accuracy metric does not express the semantic and conceptual gap between the correct and adversarial predictions, while this may be valuable information in many use-cases.

In this paper we present two novel evaluation metrics, designed to specifically address the unique challenges posed by AE attacks. 
The new metrics reflect the gap between the correct NN predictions and the adversarial predictions, i.e., the NN's prediction output when under attack, in two cases: (a) without a defense algorithm or (b) when a defense algorithm is employed.
These metrics enable us to go beyond binary metrics and compare the relative strength of attacks and defenses. 
We consider \textit{strong attacks} and \textit{weak attacks} as attack that results in a wider or narrower gap, respectively.
Similarly, \textit{strong defenses} manage to narrow that gap more than  \textit{weak defenses}. 

Since the gap is an outcome of the importance order by which the different categories are organized within the NN output vector, the suggested evaluation relates to the literature of information retrieval and ranking problems. 
The current paper concentrates on binary and multiclass classification and on the image processing domain. 
However, the presented algorithms may also be applicable to other domains in which attacks may occur, such as banking and finance~\cite{fursov2021adversarial}, cyber security~\cite{berman2019survey}, automatic speech recognition~\cite{carlini2018audio}, or medical machine learning~\cite{finlayson2019adversarial}. 

\textbf{Contributions:}
Our contributions herein are two-fold:
\begin{itemize}
    \item Novel evaluation metrics -- we present two new ranking measures, specifically utilized to accommodate classification in adversarial settings. Our metrics enable a new perspective for the comparison of attacks and defenses.
    \item Defense and attacks quality assessment --  we conduct an extensive comparison of state-of-the-art attacks and defenses. These attacks and defenses have been presented in different papers, but have not yet been compared in a rigorous manner. 
\end{itemize}

The rest of this paper proceeds as follows: in Section \ref{sec:related} we present a brief review of the relevant literature, starting with adversarial attacks, moving to defense methods against attacks, and ending with assessment methods in both deep-learning security and information retrieval. 
Then, in Section \ref{sec:method}, we discuss the required attributes for evaluation in adversarial settings, and the challenges that these settings impose.
In Sections \ref{sec:NDCG} and \ref{sec:DRR} we introduce our two new evaluation metrics.
Section \ref{sec:eval} describes our experimental setup for comparing different evaluation metrics, and the experimental results are shown and discussed in Section \ref{sec:results}.
Finally, in Section \ref{sec:conc}, we summarize the current work and its future directions.

\section{Related Work}
\label{sec:related}

\subsection{Adversarial Attacks}
A basic way to create an attack is to gradually add noise to the input, until the input is misclassified (e.g., a salt and pepper attack~\cite{chan2005salt}).
More sophisticated algorithms add noise in a more focused and effective way, enabling
the attacker to control the target adversarial category, the strength of the adversarial effect, etc. 
The differentiability of NNs allows to define a variety of optimization problems that focus on satisfying the attacker's demands with minimal added noise. 
Techniques for solving such problems include the use of simple gradients (e.g., FGSM~\cite{goodfellow2014explaining}), estimations of the Hessian (e.g., L-BFGS~\cite{szegedy2013intriguing}), or evolutionary algorithms~\cite{vidnerova2020vulnerability}, in one-shot or iterative way (e.g., DeepFool~\cite{moosavi2016deepfool}), with restricted $l_2$-, $l_\infty$-, or $l_0$-norm.

The above techniques can be applied on an ensemble of models, and thus increase the probability of creating a transferable attack, or even a universal attack~\cite{mopuri2017fast}. 
Some attacks target other aspects of NN computation; for example, they attempt to change the heatmaps produced by various interpretation methods~\cite{zhang2018interpretable,ghorbani2019interpretation}, or attack through model manipulation~\cite{heo2019fooling} or through poisoning the training data~\cite{yang2017generative,shafahi2018adversarial} rather than through input perturbations. 

\subsection{Defense methods}
The arms-race between attackers and defenders has led to the development of many defense techniques. 
Sometimes the defense strategies reflect different interpretations of the underlying causes of NN vulnerability to adversarial attacks. 
The insufficient mapping of the input space during training~\cite{tabacof2016exploring} was argued to be a consequence of the high dimensionality of the input or its non-linear processing~\cite{bhagoji2017dimensionality,gilmer2018adversarial,shafahi2018adversarial}, in which case a dimensionality reduction could help circumvent many attacks. 
Similar motivation led to different loss function or activation function modifications in order to enhance the model's robustness~\cite{papernot2016distillation,sensoy2018evidential}.
As opposed to the approaches which regard the complexity of NN computation, Goodfellow et al.~\cite{goodfellow2014explaining} put the blame on the linear summation of the initial perturbations in deep models, and suggested defending the models through adversarial training, where AEs are part of the training set. 
Many have tried to just cancel the adversarial perturbation using different kinds of input transformations, such as resizing or jpeg compression~\cite{guo2017countering,xu2018feature}. 
Others used another model or function in order to spot suspicious inputs (see~\cite{akhtar2018defense} for an extended literature review). 
Some defenses focus on masking the computational process of the model, for example through non-differentiable layers~\cite{buckman2018thermometer}. 
Ilyas et al.~\cite{ilyas2019adversarial} claim that NNs do learn to classify correctly based on their training set, and that their vulnerability reflects higher-order features that exist in the dataset and are not accessible to humans. Therefore, their suggested defense was to train on special datasets that do not contain such features. 

As new defenses emerge, adversaries adapt their attacks and create more sophisticated AEs that break many of the known defenses~\cite{athalye2018obfuscated}. 
Carlini et al.~\cite{carlini2019evaluating} claim that many defenses fail to correctly evaluate their effectiveness, either by considering only a small subset of possible attacks or by underestimating the possibility of adaptive attacks. 
Several works have therefore developed certified defenses, which provide provable robustness against attacks under certain conditions~\cite{tjeng2017evaluating,wong2018provable,lecuyer2019certified}. 
Despite the extensive research, to the best of our knowledge there is no defense that guarantees absolute robustness against AEs.  

\subsection{Evaluation Metrics}
Research on security assessment of deep learning in adversarial environment has focused mainly on available defense methods and different threat models. 
However, several papers consider the question of what assessment methodology and evaluating metrics should be used~\cite{bastani2016measuring, tjeng2017evaluating, carlini2017ground, chen2020roby}.
Batsani et al.~\cite{bastani2016measuring} used the distortion of AEs as the robustness metric, while Weng et al.~\cite{weng2018evaluating} proposed a new metric for robustness called CLEVER, based on extreme value theory. 
Peng et al.~\cite{peng2020evaluating} proposed the EDLIC framework for a quantitative analysis on different threat models and defense techniques.
They also introduced a set of evaluating metrics based on distortion level, accuracy, and the Euclidean distance between predictions. 

The accuracy of a model is usually measured using the top-1 prediction, even though sometimes the top-5 predictions are also considered. 
This metric reflects the number of times the model predicted the correct label, and it is also the basis for several commonly-used evaluation measurements~\cite{powers2020evaluation}.
The most fundamental measurements are Recall (also called Sensitivity), which stands for the proportion of real positive cases that are correctly predicted as such (i.e., true positives), and Precision (also called Confidence), which is the ratio between true positive predictions and the overall positive predictions. 
The same procedure can be applied to the negative predictions, which yields the Specificity (also called Inverse Recall) and Inverse Precision measures. 
Fallout is the proportion of real negatives in the predicted-positive cases, and Receiver Operating Characteristics (ROC) analysis plots the recall against the fallout. 
In addition, F-measure (or F1 score) is a function of Precision and Recall that can be found in deep-learning literature.

Classification metrics are mostly binary, i.e., a prediction is either correct or incorrect. 
However, we would like to evaluate the prediction \textit{ranking}. 
A domain that shares a similar objective is the evaluation of learning to rank techniques (LTR) \cite{liu2011learning}, which are mostly studied in the context of information retrieval (IR) and recommender systems. 
The goal of LTR is to rank a list of objects, e.g., a search engine is required to rank websites from most to least relevant for a given query. 

A known interface between LTR and neural network is referred to as neural ranking models~\cite{dehghani2017neural}, where NNs are used to learn a ranking function~\cite{guo2020deep}. 
Some approaches focus on learning low-dimensional representations of the input and then performing ranking using traditional IR models or other similarity metrics~\cite{salakhutdinov2009semantic,mikolov2013distributed}. 
Others attempt to directly learn the ranking function~\cite{huang2013learning,hu2014convolutional} using various network architectures and learning objectives, as reviewed in~\cite{guo2020deep}. 
Qin et al.~\cite{qin2010general} propose to directly optimize ranking metrics such as NDCG, a framework that was revisited by~\cite{bruch2019revisiting} in light of recent advances and was found empirically beneficial. 
However, as discussed in~\cite{bruch2019revisiting}, most techniques do not optimize a ranking metric directly but rather use methods such as smoothed approximation of the metric~\cite{taylor2008softrank} or an indirect boosting~\cite{wu2010adapting}. 

The above studies used NNs' output for ranking, whereas herein our goal is to apply ranking assessment to the classification output of NNs. 
Many evaluation metrics of ranked lists exist. 
One of the common ones is the mean reciprocal rank (MRR) which is simply the multiplicative inverse of the rank of the first correct answer. 
Normalized discounted cumulative gain (NDCG) \cite{jarvelin2017ir} is another known metric, which we will hereby expand. 
Other common metrics such as  expected reciprocal rank (ERR) \cite{chapelle2009expected} consider the user's satisfaction from a ranked list and are irrelevant in our context. 

To conclude, evaluation for neural ranking models exists, as described above. Evaluation for classification models also exists but is essentially binary. 
Yet, while in classification tasks NNs output a ranking of the classes in question, it is common in these tasks to consider only the result ranked in the first place, and ignore the rest of the ranking.
We suggest herein to evaluate the whole ranking and thus provide a more accurate evaluation for classification tasks.

\section{Methodology}
\label{sec:method}

When the top-most prediction is correct for a given AE, it is common to consider this as a complete defense against the AE attack. Similarly, a defense is considered incomplete when the top-most prediction is wrong. This dichotomous division to complete and incomplete defenses  
does not consider other subtle scenarios. 
For example, the top-most prediction may be correct but the AE has caused: (a) a lower probability for the top-most prediction or (b) a distortion in other places in the ranked list of outputs.

Thus, we propose to increase the expressiveness of the evaluation and distinguish between various cases of complete or incomplete defenses. 
Similarly, we would like to compare different successful attacks, where the top-most prediction is always wrong, but the deviation from the true label may vary. 
Our goal is to step beyond the mere indication of whether the topmost item is classified correctly; 
instead, we evaluate the performance of the entire NN's ranked list of prediction. 
As an example, consider the image of `toy poodle' (see Table \ref{table:NDCG_examples}).  The image was modified using two different attacks -- FGSM attack~\cite{goodfellow2014explaining} and Targeted CW attack~\cite{carlini2017adversarial}. 
When the NN is not under attack, the top-5 predictions for the toy poodle image are: {`toy poodle', `miniature poodle', `standard poodle', `cocker spaniel', `seat belt'}. We herein refer to these predictions as the benign predictions. 

The FGSM attack confused the NN, causing it to classify the image as a `miniature poodle', while the Targeted CW attack caused it to classify the image as a `snowmobile'. 
Both classifications are wrong, but a `miniature poodle' is not as wrong a classification as a `snowmobile'. 
However, dichotomous classification evaluation metrics will treat both attacks as equally harmful. 
Our desired metric should be able to account for the position of the predicted label in the benign ranked prediction. 
In our example, since a miniature poodle is higher than a snowmobile in the benign prediction, we would like to be able to give a lower score to the quality of the FGSM attack. 

Evaluating the quality of an outputted ranking is not a trivial problem, many metrics have been considered and there is no single metric which is optimal for every type of ranking problem~\cite{croft2010search}.
When dealing with attacks, it is necessary to consider two unique problem features: the baseline ranking is unknown, and the ranking is nonlinear.

\smallskip 
\textbf{The baseline ranking is unknown.} 
In order to compare different classification decisions under attacks or defenses, the top-1 prediction in each case should be evaluated by its distance from the true top-1 prediction.
Natural language processing techniques such as word2vec~\cite{mikolov2013distributed} might provide such distance estimation. 
In this method, a NN is used to create word embeddings, where each distinct word in the vocabulary is represented by a unique vector, and the semantic relationships between words are expressed and measured as the distances in the vector space. 
However, this universal and fixed proximity function between categories will not necessarily allow an accurate estimation of the adversarial effect, because the associations between the categories are context-dependant.
Some categories relate to each other,
e.g., different kinds of small dogs are closely categorized using word2vec, truly reflecting the relationship between these categories.  
In other cases, word2vec may fail when the unique features of a given image may effect the relationships between the categories. 
For example, a specific posture of the dog, the image perspective, or the background texture may increase its similarity to other objects that are not close according to word2vec. 
A poodle against a background of flowers may result in high ranking for objects from the plants categories, even though the word embedding of `poodle' and `plants' are not close in the vector space.
In our example, for the `toy poodle' image, the benign NN ranked `seat belt' in the fifth place, whereas word2vec does not consider a `toy poodle' and a `seat belt` as close categories. 
     
\smallskip    
\textbf{The ranking is not linear.}
Confusion between top-ranked categories (e.g., replacing `toy poodle' with `miniature poodle') is much more probable than replacement of a first-ranked category with a distant one. 
Therefore, the effort needed by an attack to cause a considerable deviation from the original category is not a linear function of the distance between categories on the scale. 
Back to the `toy poodle' example (Table \ref{table:NDCG_examples}), in the benign setting the fourth prediction of the model is `cocker spaniel', while the fifth prediction is `seat belt'. 
This is a dramatic decrease in relevance, and an attack that manages to cause misclassification of the image as a seat belt rather than some kind of dog is considered much stronger. 
However, an attack that causes a confusion between the 49th and 50th ranked categories should not be considered as effective to the same extent. 
Therefore, a metric such as correlation, which treats all mismatches uniformly, is less suitable.

\smallskip
To this end, in the following sections we propose two novel evaluation metrics. The first, introduced in Section~\ref{sec:NDCG}, mainly focuses on the ranking of categories and is based on the $\mathit{NDCG}$ measure~\cite{jarvelin2017ir}. The second evaluation metric, presented in Section~\ref{sec:DRR}, relates additionally to the relevance scores from which the ranking is derived, and is based on the MRR measure~\cite{chapelle2009expected}.
The choice to consider two distinct metrics follows the fact that there are two main approaches to assess ranked lists in the IR literature.
However, the two approaches also reflect a deeper difference in the core logic guiding the assessment.
The $\mathit{NDCG}$-based metric focuses on quantifying how wrong the top-$k$ predictions are.
Each such prediction is punished in proportion to its deviation from the benign input prediction. 
On the other hand, the score of the MRR-based metric is only determined by the rank and confidence level of the correct top-1 category within the adversarial prediction list. 
In this sense, it is less concerned with how wrong the predictions are, and only focuses on how right the ranking of the true category is within the adversarial prediction.
Hence, these novel metrics offer enhanced and diverse evaluation capabilities that can be utilized in various scenarios.

\section{$\mathit{NDCG}$-based Evaluation Metric}
\label{sec:NDCG}

Our first metric is based on Normalized Discounted Cumulative Gain ($\mathit{NDCG}$)~\cite{jarvelin2017ir}.
We first describe $\mathit{NDCG}$ and then continue to describe our metric. 
 $\mathit{NDCG}$ was originally proposed in the IR domain for evaluation of the ranked list of results that a search engine outputs for a given query.
 It assumes that correct ranks at the top of the result list are more valuable than correct ranks at the bottom of the list.
 Thus, the evaluation score considers the relevance and the position of each element in the list.
 First, the Discounted Cumulative Gain (DCG) is computed.
 The relevance of each item is reduced as a function of its location in the list, so that the gain of later items is progressively discounted.
 This reflects the fact that earlier search results are more likely to be explored than the recommendations at the bottom of the list, and are therefore more important. DCG is computed as
\begin{equation}
\text{DCG}_k = \sum_{i=1}^k \frac{2^{rel{_i}}-1}{\log (1+i)}
\label{eq1}
\end{equation}
where the parameter $k$ is the number of the first items to be considered and the relevance $rel_i$ is the predefined relevance of item $i$, regardless of its location in the list.
After the initial relevance of the $k$ items are discounted by their position and accumulated, the DCG score is normalized as

\begin{equation}
\text{NDCG}_k =  \frac{\text{DCG}_k}{\text{IDCG}_k}
\label{eq2}
\end{equation}
where IDCG is the ideally-ordered list.
For example, if a search engine outputs a list of 5 results, then the relevance score of each of these recommendations is given by their measured affinity to the original query.
Assuming a ranking between 0-3, a ranked recommendation list could look like [2, 3, 3, 0, 1], whereas a perfect ordering of those recommended items would be [3, 3, 2, 1, 0].
The global $\mathit{NDCG}$ score can also be itemized, where each item's discounted gain is divided by its ideal counterpart, i.e., the discounted gain of the item at the same location in the perfect list.   
Note, therefore, that $\mathit{NDCG}$ evaluates the ranked list of results, regardless of the numerical value of each result. 

Similarly to recommendation systems, the output of NNs also consists of ordered lists of predicted classes, and our objective is to evaluate attacks and defenses based on their effect on the order of the model's predictions.
Therefore, customizing the $\mathit{NDCG}$ measure requires the following considerations:
\begin{enumerate}
    \item defining relevance scores
    \item setting the value of $k$
    \item defining ideal ranking for normalization
\end{enumerate}

In the remainder of this section the evaluation metric will be demonstrated on the evaluation of attacks, but it can be applied similarly to evaluation of defenses.

\subsection{Relevance score}
Naturally, the relevance scoring of a model's output should relate to the confidence of that model in its predictions. We propose to use the predictions of the model for a benign input as a source for relevance scores and as an ideal grading, by which attack and defense results should be normalized. The underlying assumption is that a well-trained model is not only capable of providing the correct top-1 prediction, but also of capturing the relationships between different categories. This ability is the essence of ``learning'', which enables the model to generalize beyond the training data. Therefore, the entire prediction list of the model when given a benign input is considered here as a ground truth to determine the relevance scores. 

\subsection{Setting k}
Any specific input results in a unique distribution of confidence among the entire list of output categories. Thus, the parameter $k$, which controls the length of the list used for evaluation, should be determined dynamically, according to the specific characteristics of each individual example. 
In the case of ImageNet classification, though there are $n=1,000$ different output categories, the confidence of the model is usually distributed among a very small portion of the list, while assigning infinitesimal scores to most categories. Consequently, despite its potential length, the list of relevant predictions is usually rather short.

Yet, there is a mechanism within NNs that comes into action -- at the last layer of the NN a \textit{softmax} function is applied to the predictions list.
This function assigns more weight to an even narrower range of categories, while the weight of all other categories decays exponentially.
We, therefore, apply relevance scores using the normalized pre-softmax output of the model in order to reflect the original confidence relations between the categories, but determine $k$ according to the post-softmax distribution of the confidence in order to preserve only the most relevant predictions as part of the scored list.
Specifically, a set of post-softmax predictions
$\{p_1^{(e_b)},p_2^{(e_b)},\ldots,p_n^{(e_b)}\}$  for a specific benign example $e_b$ is scored using the corresponding pre-softmax values  
$\{l_1^{(e_b)},l_2^{(e_b)},\ldots,l_n^{(e_b)}\}$
in the following way:
\begin{equation}
R_{p_i}^{(e_b)} =
\begin{cases}
    \frac{l_i^{(e_b)}}{\sum_j l_j^{(e_b)}}             & i\leq k_1\\
    0                       & \text{otherwise}
\end{cases}
\label{eq3}
\end{equation}
The post-softmax predictions
$\{p_1^{(e_b)},p_2^{(e_b)},\ldots,p_n^{(e_b)}\}$ are arranged in descending order, and $k_1=\displaystyle\max_j p_j^{(e_b)}\geq \gamma_b$, where $\gamma_b =[0,1]$ is the prediction threshold parameter for benign inputs.
For example, for $\gamma_b = 0.01$, only the $k$ highest post-softmax predictions that are above this 0.01 threshold will receive relevance scores in proportion to their original pre-softmax value, and all the other $n-k$ predictions will be zeros.
In the toy-poodle example in Table~\ref{table:NDCG_examples} (benign column) the predictions are: $\{0.62, 0.37, 0.001, \ldots \} $. 
Setting $\gamma_b = 0.01$ results in $k_1 = 2$, and therefore only the two top-most predictions receive relevance scores. The benign relevance values are $\{0.504, 0.495, 0, \ldots, 0\}$. They are computed as follows: first, the pre-softmax values of $a_b$ are sorted in descending order: $\{20.328, 19.829, 14.286, \ldots, -6.023\}$; then, they are rescaled into the range $[0,1]$, resulting in values $\{1, 0.98, 0.77, \ldots, 0\}$. Since $k_1 = 2$, the only two top-most values are considered and normalized, in order to receive their relative contributions. 

For an adversarial example $e_a$ that was created on the basis of $e_b$, the relevance of the predictions is determined as follows:
\begin{equation}
R_{p_i}^{(e_a)} =
\begin{cases}
    R_{p_{i_{match}}}^{(e_b)} & i\leq k_2\\
    0                       & \text{otherwise}
\end{cases}
\label{eq4}
\end{equation}
where $k_2=\displaystyle\max_j p_j^{(e_a)}\geq \gamma_a$ with the threshold parameter for adversarial inputs: $\gamma_a= [0,1]$
and
$p_{i_{match}}$ is the prediction within the $e_b$ output that corresponds to the category of $p_i$ (note that this prediction is not necessarily located in the $i$th place). In our running example for the FGSM attack $e_a1$,
$p_1 = 0.566$ (`miniature poodle'),
$p_{1_{match}} = 0.37$ (this is the prediction value assigned to `miniature poodle' in the benign case), and
$R_{p_1}^{(e_a)} = R_{p_{1_{match}}}^{(e_b)} = 0.495$.

Note that $\gamma_a$ differs from $\gamma_b$, and should usually be lower, since the original prediction values in the adversarial predictions list are commonly substantially lower. 
In the toy-poodle example of Table~\ref{table:NDCG_examples}, setting $\gamma_a = 0.001$ results in $k_2 = 613$, and therefore all top-most 613 predictions receive the corresponding relevance scores of their benign counterparts.
However, since $k_1 = 2$, only two of those relevance scores are non-zeros. 

The number $k$ of non-zero items within the predictions lists reflects the confidence level of the model for both the AE and the original seed.
Suppose that during an attack, the original confidence distribution is disturbed and is now spread over more categories, then the AE scored predictions-list can potentially assign relevance values to more categories.
Nevertheless, if the attack caused the model to assign higher values to irrelevant categories, for example assigning the benign image `toy poodle' with the categories: `shark' `broccoli'
to be uncertain whether the input image is a shark, a broccoli or a toy poodle, the extra values will still be zeros, since they receive the values of their benign counterparts.
In the extreme case where an attack causes the model to predict entirely wrong categories with high confidence, the scored adversarial predictions-list will be composed of zeros.  

The discounted cumulative gain of the predictions for an image $i$ is given by:
\begin{equation}
\text{DCG}_k^{(e)} = \sum_{i=1}^k \frac{2^{R_{p_i}^{(e)}}-1}{\log (1+i)}
\label{eq5}
\end{equation}
Therefore, the more an attack 'pushes' a correct prediction down the predictions hierarchy, the more discounted is the gain of that prediction.
See, for example, the difference between the DCG scores of FGSM attack and CW attack in Table~\ref{table:NDCG_examples}, where in the later case the scores are significantly more discounted due to their location in the ordered predictions-list.
In other words, the DCG score 'punishes' predictions in proportion to their deviation from the ground-truth distribution of the benign input's predictions.

\subsection{Ideal ranking for normalization}

We can now calculate the normalized $\text{DCG}_k$ for an adversarial example $e_a$. 
Following the logic of Equation \ref{eq2} we obtain:
\begin{equation}
\text{NDCG}_{k_2}^{(e_a)} =  \frac{\text{DCG}_{k_2}^{(e_a)}}{\text{DCG}_{k_1}^{(e_b)}}
\label{eq6}
\end{equation}
As a consequence, the $\mathit{NDCG}$ score of benign examples is always one, and attacks are considered more effective as this score decreases (accordingly, an effective defense increases the score).
Another advantage of $\mathit{NDCG}$ is that this metric assigns a score to each of the output categories. 
These scores can reflect different attack trends; if an attack only changes the top-1 category, then there is only a local drop in the score of the first prediction while the rest of the prediction-list's scores saturate.
Alternatively, if an attack changes the entire semantic neighborhood of all the top predictions, then the decrease is significant across all the predicted classes. 
The last row of Table~\ref{table:NDCG_examples} presents examples of these two scenarios, where for the FGSM attack there is only a local decrease in the scores of the two first-ranked categories, while for the CW attack the initial decrease is much more dramatic and the increase of that score along the next categories is more gradual.

\newcolumntype{C}{>{\centering\arraybackslash}X}
\begin{table*}
\setlength\tabcolsep{2pt}%
\begin{tabularx}{\textwidth}{@{}c*{4}{C}@{}}

    \textbf{Attack type} & Benign (no attack) & FGSM Attack & Targeted CW Attack \\
    \hline
\textbf{Input image} &
   \includegraphics[align=c, width=0.6\linewidth, height=\linewidth, keepaspectratio]{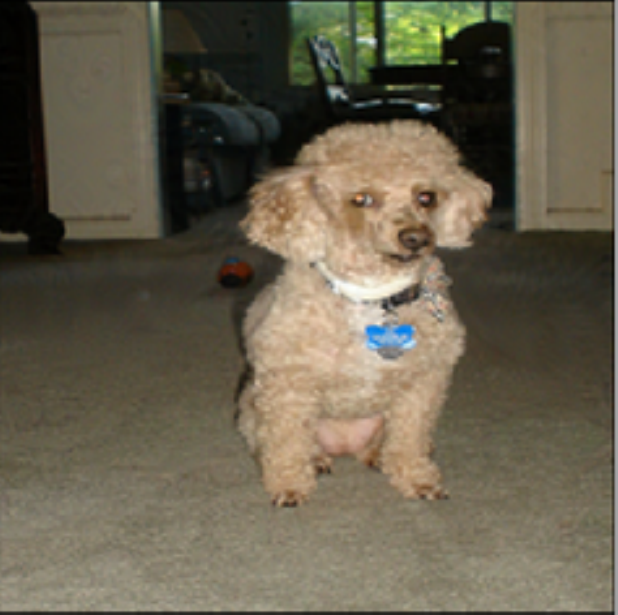} &
   \includegraphics[align=c, width=0.6\linewidth, height=\linewidth, keepaspectratio]{toy_poodle.pdf} &
   \includegraphics[align=c, width=0.6\linewidth, height=\linewidth, keepaspectratio]{toy_poodle.pdf} \\
\textbf{} & $e_b$ & $e_{a1}$ & $e_{a2}$ \\
\hline\\
\textbf{Top-5 predictions} & \makecell{`toy poodle', 0.62 \\ `miniature poodle', 0.37 \\`standard poodle', 0.001\\`cocker spaniel', 0.0004\\`seat belt', 3.69e-05\\} & \makecell{`miniature poodle', 0.566\\`toy poodle', 0.428\\`standard poodle', 0.004\\`cocker spaniel', 0.0004\\`Irish water spaniel', 6.42e-05\\} & \makecell{`snowmobile', 0.005\\`Ches. Bay retriever', 0.005\\`hyena', 0.005\\`miniature poodle', 0.005\\`jeep', 0.005\\}\\ \hline\\
\textbf{k} & $k_1 = 2$ & $k_2 = 2$ & $k_2 = 613$ \\
\hline\\
\textbf{Top-12 relevance scores} & \makecell{0.505\phantom{0} \hfill 0.495\phantom{0} \hfill  0.\phantom{0000}  \hfill  0.\\0.\phantom{0000} \hfill 0.\phantom{0000} \hfill  0.\phantom{0000} \hfill 0. \\0.\phantom{0000} \hfill   0.\phantom{0000} \hfill  0.\phantom{0000} \hfill  0.} & \makecell{0.495\phantom{0} \hfill 0.505\phantom{0} \hfill  0.\phantom{0000}  \hfill  0.\\0.\phantom{0000} \hfill 0.\phantom{0000} \hfill  0.\phantom{0000} \hfill 0. \\0.\phantom{0000} \hfill   0.\phantom{0000} \hfill  0.\phantom{0000} \hfill  0.} & \makecell{0.\phantom{000}  \hfill 0.\phantom{000}  \hfill 0.\phantom{0}  \hfill   0.495\\0.\phantom{000}  \hfill 0.\phantom{000} \hfill 0.\phantom{0000}  \hfill 0.\\ 0. \phantom{000}  \hfill 0.\phantom{000}  \hfill 0.505 \phantom{0}  \hfill 0.}\\
\hline\\
\textbf{Top-12 DCG scores} & \makecell{0.419  0.677  0.677  0.677 \\ 0.677  0.677  0.677  0.677 \\ 0.677  0.677  0.677  0.677
} & \makecell{0.409  0.674  0.674  0.674 \\ 0.674  0.674  0.674  0.674 \\ 0.674  0.674  0.674  0.674
} & \makecell{0.\phantom{000} 0.\phantom{000} 0.\phantom{000} 0.176\\ 0.176  0.176  0.176  0.176\\ 0.176  0.176  0.293  0.293}\\
\hline\\
\textbf{Top-12 NDCG scores} & \makecell{1.\phantom{0000} \hfill 1.\phantom{0000} \hfill 1.\phantom{0000} \hfill 1.\\ 1.\phantom{0000} \hfill 1.\phantom{0000} \hfill 1.\phantom{0000} \hfill 1.\\ 1.\phantom{0000} \hfill 1.\phantom{0000} \hfill 1.\phantom{0000} \hfill 1.} & \makecell{0.977 0.995 0.995 0.995\\ 0.995 0.995 0.995 0.995\\0.995 0.995 0.995 0.995} & \makecell{0.\phantom{000} \hfill 0.\phantom{000} \hfill 0.\phantom{000} \hfill 0.260\\0.260  0.260  0.260  0.260\\0.260  0.260  0.433  0.433}\\
\hline\\

\end{tabularx}
\caption{$\mathit{NDCG}$ examples of benign input and two different attacks, where $\gamma_b$ = 0.01 and $\gamma_a$ = 0.001. It can be observed that the CW attack has a significantly stronger effect on the model's predictions, as reflected in the higher value of $k_2$ and lower $\mathit{NDCG}$ scores. Contrary to that, the FGSM attack mainly switches between the top 2 predictions, leaving the rest of the predictions almost the same, as reflected in the minor $\mathit{NDCG}$ drop at the first prediction.} \label{table:NDCG_examples}
\end{table*}

\section{Defense Reciprocal Rank}
\label{sec:DRR}
This measure is based on the reciprocal rank measure~\cite{chapelle2009expected}, which is commonly used in information retrieval literature.
Reciprocal rank calculates the multiplicative inverse of the rank position of a category within a given ordered list. 
If the benign category appears at the top-1 position, the reciprocal rank will be 1.
If the benign category is at rank 2 then the reciprocal rank is $1/2$, for rank 3 it is $1/3$ and so on. The motivation behind this approach is simple -- in an ordered list of predictions, the further away the benign category is ranked, the lower the score it receives. 
In our case, given an input image, the NN outputs a probability distribution over the output categories.
When sorted in a descending order, we can compute the reciprocal rank at which the benign category is predicted (i.e., the rank where the correct category label of the input image appears).
As the output list can be very long, and therefore may result in negligible reciprocal rank scores, we restrict the measure to only consider the first top-$k$ results of the ordered predictions list.
This way, only if the true category is predicted within the top-$k$ predictions, the score is proportional to its proximity to the top-$1$ location, and otherwise the reciprocal rank is set to $0$. 

We modified the reciprocal rank, so that it also considers the confidence of the model in its prediction. 
The intuition behind this is that a correct classification of an AE with higher confidence is better than a correct classification with lower confidence.
In this case, the lower confidence may indicate that the attack, though failed to fully deceive the NN, gained some effect on the NN 
so that it is can now be considered as more vulnerable. 
However, if the NN misclassifies an AE, then the confidence in the benign prediction should only improve the score within the bounds of the rank of that prediction.
Namely, if the benign category of `toy poodle' is now located in rank $2$, the score cannot surpass the score of `toy poodle' predicted at rank $1$, regardless of the confidence level of the predictions.

Assuming all predictions are given as probabilities (i.e., sum up to one), we note that the possible range of confidence level is narrowed as the rank increases.
The rank $1$ prediction's confidence may vary between zero and one. 
However, the rank $2$ category can only be at confidence between zero and half, while rank $3$ prediction can only be between zero and third, etc.
Therefore, we normalize the confidence of the prediction at each rank $i$ to be between the maximal value of rank $i-1$ and the maximum value of rank $i$.
The defense reciprocal rank ($\mathit{DRR}$) score is thus given as:
\begin{equation}
DRR_k =
\begin{cases}
    \frac{\bar{P}_i}{R_i + 1} + \frac{1}{R_i + 1} & i < k\\
    0 & \text{otherwise}
\end{cases}
\label{eq7}
\end{equation}
where $\bar{P}_i$ is the probability assigned to the benign category $i$, and $R_i$ is the rank of that category within the ordered predictions list. This way, the first term of the $\mathit{DRR}$ equetion squeezes the original $\bar{P}_i$ value between zero and $1 / (R_i + 1)$, while the second term ensures that the lower bound of the new range will be $1 / (R_i + 1)$.

Back to the toy-poodle example, in the NN's output to $e_{a1}$ the benign category `toy-poodle' is located in the second place (rank $2$) with probability $0.428$.
Therefore the $\mathit{DRR}_5$ score is calculated as $0.428/3 + 1/3 = 0.476$.
The $\mathit{DRR}_5$ of $e_{a2}$, however, is $0$ since the benign category is not one of the top-5 predictions.   

Usually, the probability distribution over the output categories is given as a softmax function.
This nonlinear mapping is easier to interpret as a classification decision, since it is closer to the argmax function while still remaining differentiable.
However, to capture the relationships between the different categories, it is more suitable for our needs to use a linear normalization of the logits, i.e. the pre-softmax layer of the network.
Though the results in this paper will only consist of this linear logits normalization, similar results are obtained when using the standard softmax probabilities as well.

\section{Experimental Evaluation}
\label{sec:eval}

In our experiments, we applied the suggested new metrics ($\mathit{NDCG}$ and $\mathit{DRR}$) to several attacks and defenses, and examined 
how the new scores reflect their diversity. 
For example, even when considering only successful attacks, which cause the NN to misclassify, the observed effect may vary between different attack types.
In general, it is expected that targeted attacks would have deeper and more substantial effect than the untargeted attacks, since they are not satisfied by merely misclassifying with the most approachable wrong category. 
Similarly, iterative attacks which are more computationally demanding are expected to be, on average, stronger than one-shot attacks which cannot refine the adversarial perturbations.
While the traditional top-1 accuracy metric assigns identical score to all these attacks, our suggested metrics enable a distinction between different degrees of misclassification.
\subsection{Experimental Setting}
The experiments were performed on the ImageNet validation dataset~\cite{deng2009imagenet} using the pretrained VGG19 model~\cite{simonyan2014very}.
The attacks were created using the Foolbox toolbox~\cite{rauber2017foolbox}, and consist of $l_2$-bounded targeted L-BFGS and Carlini-Wagner (CW) attacks, as well as untargeted FGSM, boundary, and CW attacks. 
In order to investigate the net effect of the attacks, benign images which were misclassified by the model have been removed from the dataset. 
Similarly, failed attacks which did not cause the desired misclassification were not included in the attacks dataset.
Therefore, the attacks dataset only includes benign images, which were correctly classified by the model, and their adversarial counterparts, which cause the desired misclassification.
An experimental group includes an average of 550 pairs of benign and adversarial examples for a specific attack.
We created a total of five experimental groups, one for each of the attacks mentioned above.

We also tested the effectiveness of defenses and considered three representative methods.
The defenses have been applied to NN models under the above attacks in order to test the recovery of their original predictions.
The first defense was adversarial training, which, as the name suggests, is based on retraining the NN using adversarial examples with the true labels.
This way, the NN learns a more robust representation of the training set, by reducing the unmapped areas of its latent space.
This method was found effective even against adaptive attacks, in which the attacker is aware of the defense method and intentionally tries to break it~\cite{athalye2018obfuscated}.
We retrained the base VGG19 model for 10 epochs, following the method proposed by Goodfellow et. al.~\cite{goodfellow2014explaining}.
The second defense falls under the category of input transformation; in this approach, the adversarial perturbations are directly cancelled by defensive perturbations. 
Specifically, we applied bit-depth squeezing to the AEs before inputting them to the NN.
The squeezing was conducted under two experimental conditions, 4-bit depth and 5-bit depth, in order to examine the effect of the depth on the defense within the depth range recommended by Xu, Evans, and Qi \cite{xu2018feature}.
The third defense strategy suggests to inject randomness into the NN computational process \cite{carlini2017adversarial}. We implemented this approach via dropout of either 0.1 or 0.3 at the first fully-connected layer of the NN.
The effectiveness of the different defenses was tested under two conditions; when used against targeted attacks (CW and LBFGS) or against untargeted attacks (CW, boundary, and FGSM). 

\begin{figure}[h!]
\centering
\includegraphics[width=0.53\textwidth]{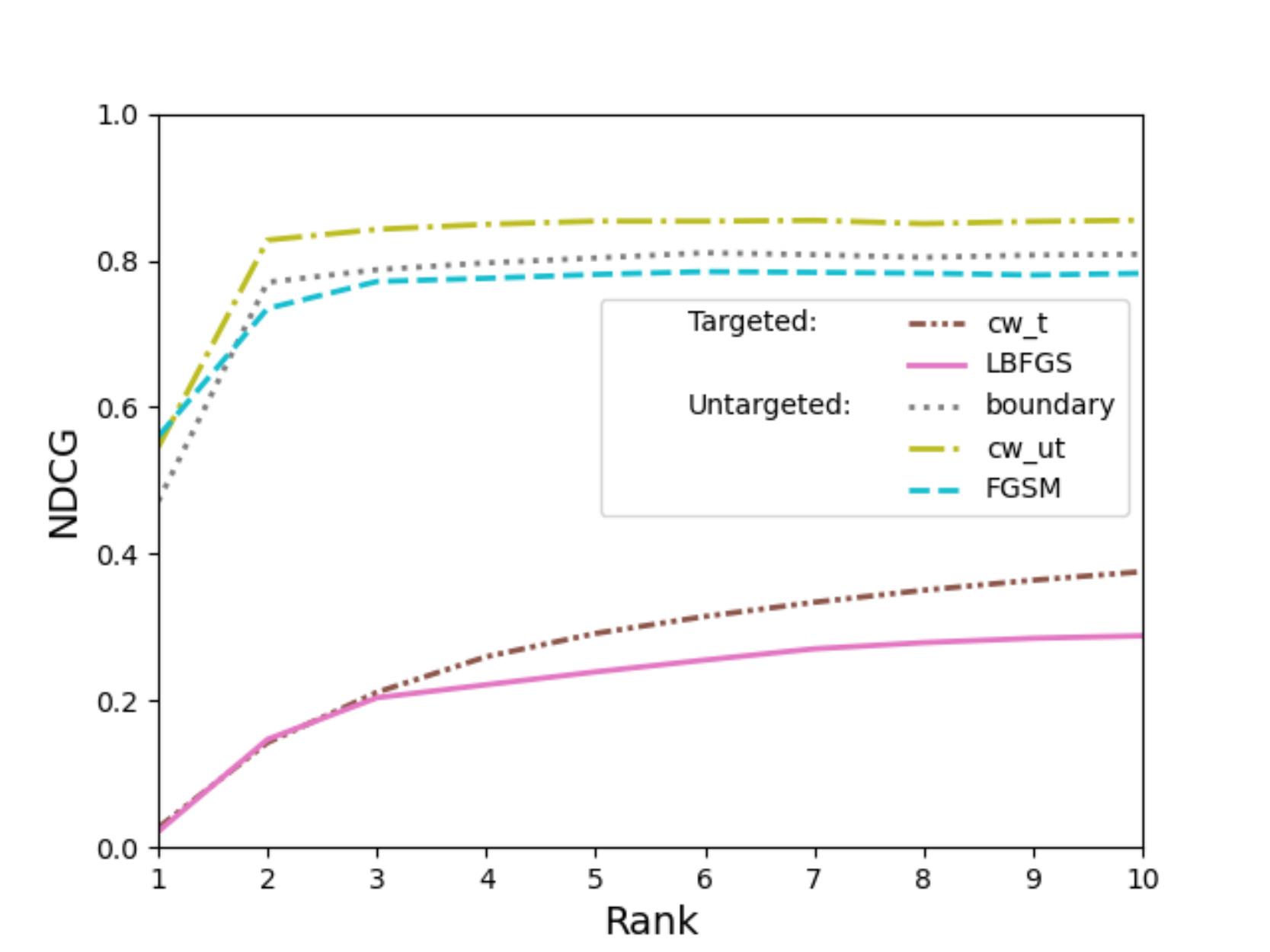}
\caption{$\mathit{NDCG}$ scores of the top-10 predictions for different attacks}
\label{fig:NDCG}
\end{figure}

Figure~\ref{fig:NDCG} presents the $\mathit{NDCG}$ scores for the ten top-most predictions, under five different attacks. 
It can be observed that there is a significant gap between the scores of the targeted and untargeted attacks.
This gap indicates that targeted attacks are stronger;
they cause top predictions to deviate from predictions ranked in close proximity to the benign top predictions. 
The effect of the untargeted attacks is mainly focused on the top-1 prediction while the following predictions saturate around 0.7 $\mathit{NDCG}$ score.
On the other hand, the targeted attacks manage to significantly reduce the score of the entire predictions list, with a continuous, yet moderate, recovery as the rank increases.
This reflects the fact that untargeted attacks are weaker and tend to keep the top predictions within their original semantic neighborhood.
Usually these attacks only change the order of the top-most prediction, or even only switch between the first and second predictions.

While observing the trend of the $\mathit{NDCG}$ scores of all top-$k$ predictions gives a better understanding of the adversarial effect, the most significant effect is found in the top-1 prediction.
Moreover, since for most applications the main concern is the classification decision of the NN as reflected in the top-1 prediction and the rest of the predictions list is of less interest, in the following experiments we will only consider the $\mathit{NDCG}$ score of the top-most rank.

In the next experiment we examine the metrics' ability to detect:
\begin{itemize}
    \item Attack-attack separation -- the distinction between attacks in terms of their quality, e.g., strong and weak attacks.
    \item Attack-benign separation -- the distinction between benign inputs and adversarial inputs.
\end{itemize}

\begin{figure}[t]
\centering
\includegraphics[width=0.45\textwidth]{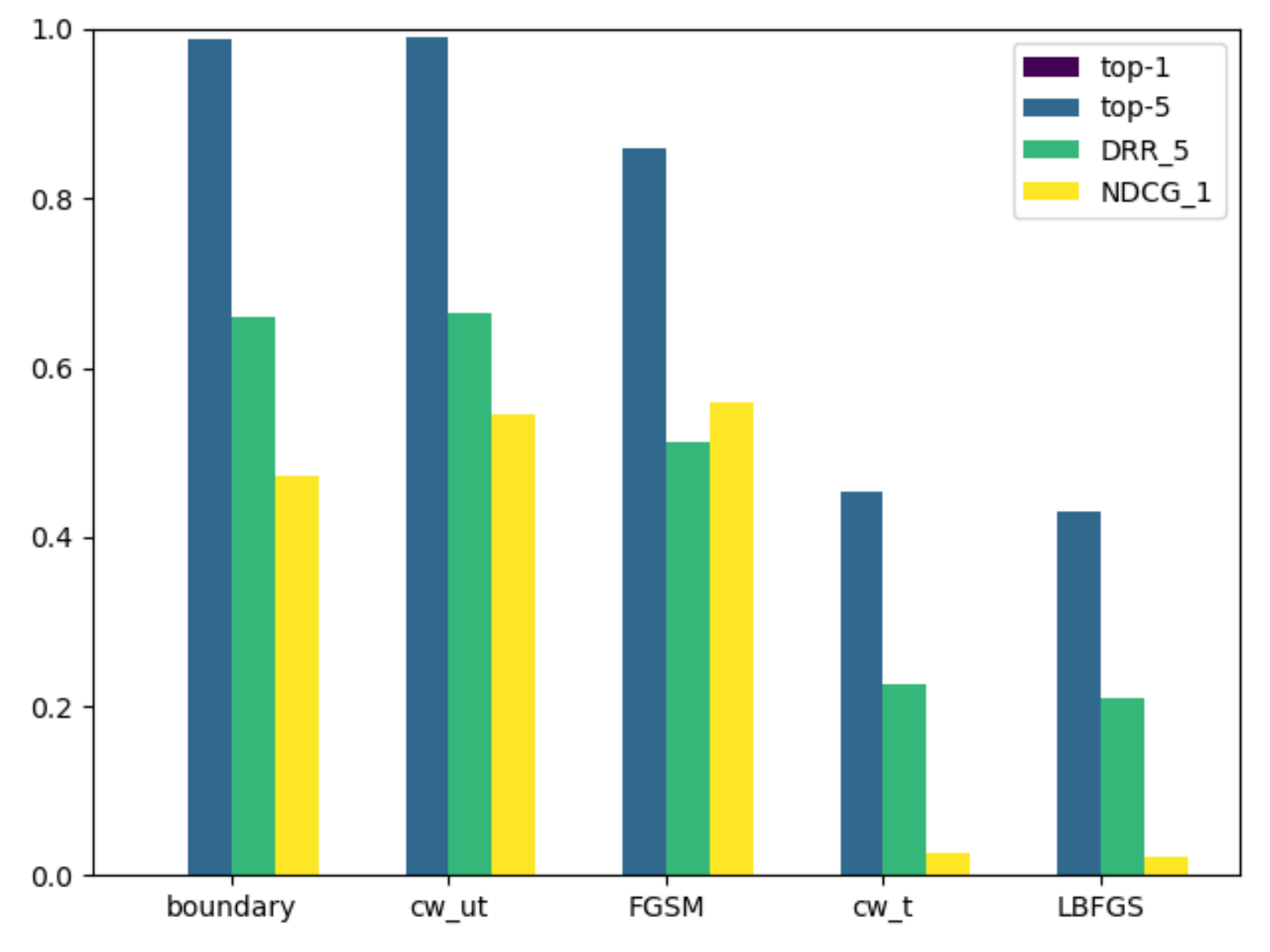}
\caption{Evaluation of attacks using different metrics}
\label{fig:attacks}
\end{figure}

\subsection{Results}
\label{sec:results}

Figure~\ref{fig:attacks} compares between different evaluation metrics of the above attacks.
The top-$1$ and top-$5$ metrics illuminate a trade-off between attack-attack separation and attack-benign separation. 
The top-$1$ metric perfectly captures attack-benign separation because the dataset consists only of successful attacks. However, it fails in attack-attack separation.
Another view of the trade-off is witnessed in the top-$5$ metric -- it nicely captures attack-attack separation, however it fails in attack-benign separation for weak attacks. 
This begs the question of how other top-$k$ metrics (such as top-2 or top-3) would perform.
They may display a better trade-off between attack-attack separation and attack-benign separation, however this will require consistent manual tuning of $k$, depending on the specific characteristics of the tested attacks.
Another shortcoming of all top-$k$ measures is that they are binary, and are therefore inherently less informative.

Our two suggested evaluation metrics successfully balance the trade-off between attack-attack separation and attack-benign separation, as they both differentiate attacks from benign inputs and also differentiate between attack types. 
For $\mathit{DRR}$, we set $k=5$ following the top-$5$ measure in the literature as an alternative to the top-$1$ measure.
This parameter seems to satisfy the demands we set for the new desired metric, though a smaller $k$ will produce stricter evaluation and lower scores, similar to the $\mathit{NDCG}_1$ measure. 
Overall, the $\mathit{DRR}_5$ measure can correctly identify cases in which the true category of the input is present as one of the top-$5$ predictions. This is more common in weak attacks that cause the NN a minor miss-classification. 
The $\mathit{NDCG}_1$ measure, on the other hand, does not consider the rank of the true category but rather the similarity between the actual top-$1$ prediction and the true prediction.
The gap between the $\mathit{NDCG}_1$ scores and the $\mathit{DRR}_5$ scores indicates that in many cases, the top-$1$ prediction can be wrong, or even very wrong, but the true category may still be found as one of the top-$5$ predictions as well. 

\begin{figure}[t]
\centering
\includegraphics[width=0.45\textwidth]{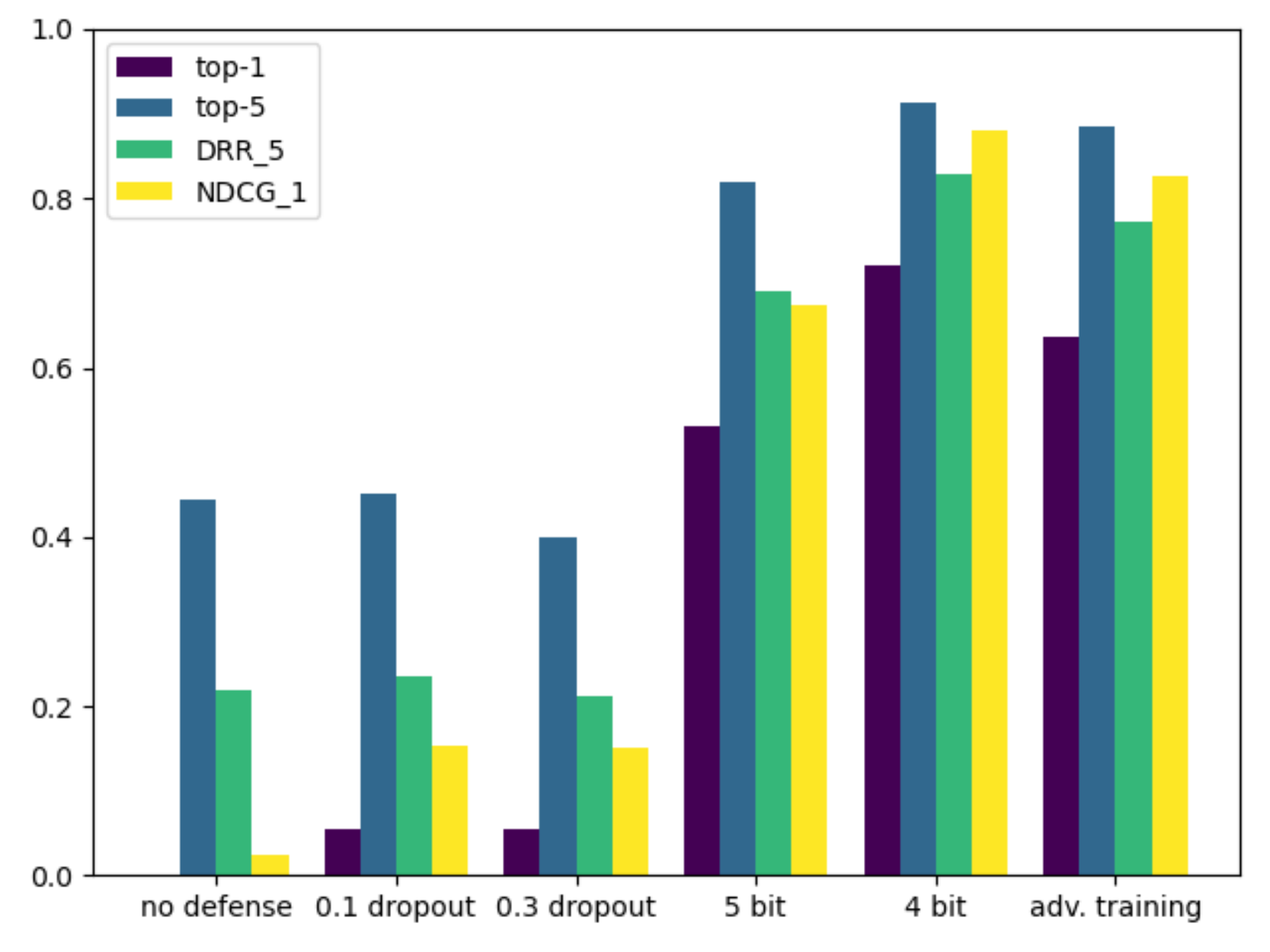}
\caption{Evaluation of defenses against targeted attacks using different metrics}
\label{fig:targeted_defenses}
\end{figure}

\begin{figure}[t]
\centering
\includegraphics[width=0.45\textwidth]{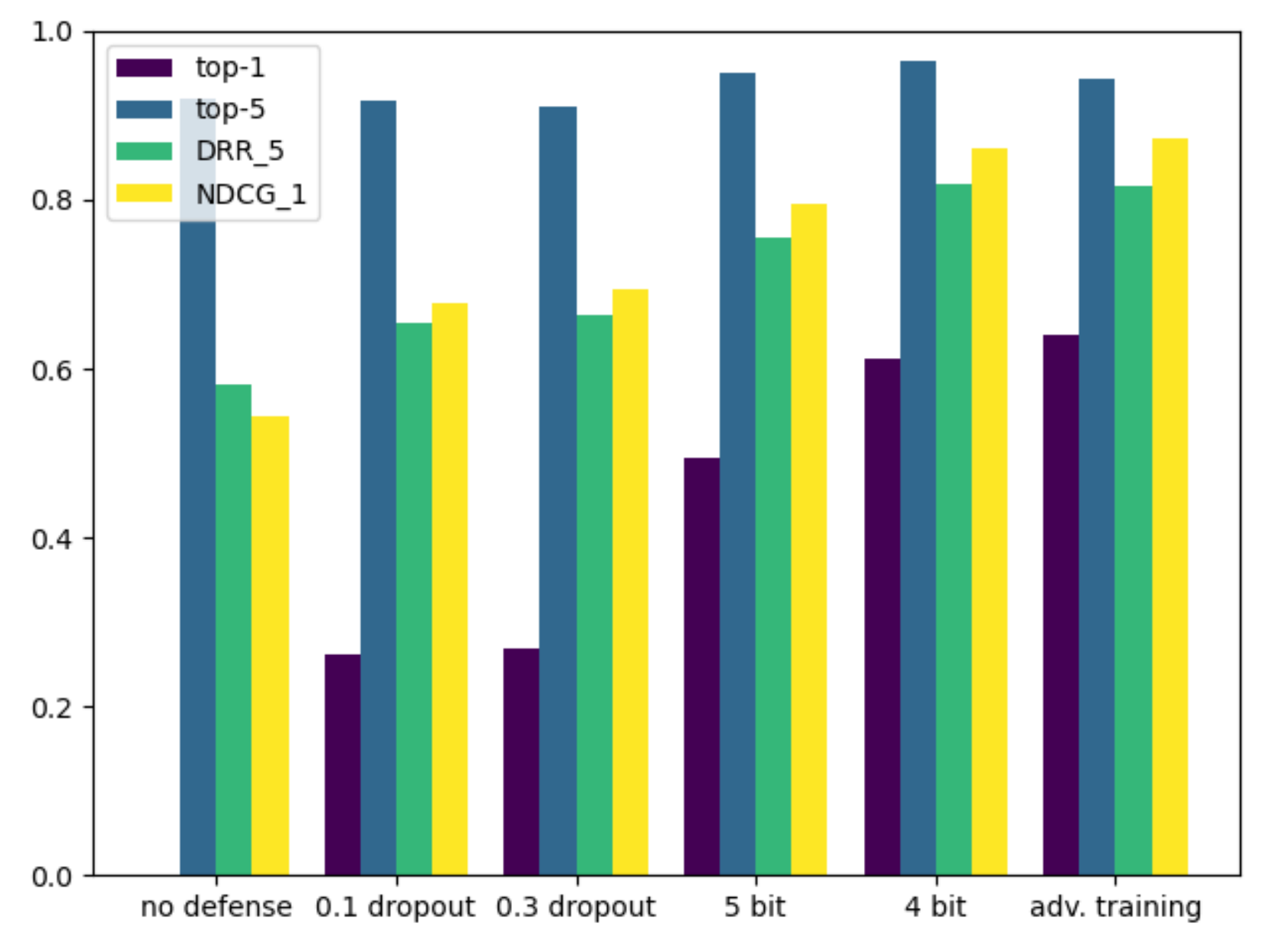}
\caption{Evaluation of defenses against untargeted attacks using different metrics}
\label{fig:untargeted_defenses}
\end{figure}

Applying different defenses to the attacked NN demands a more complicated evaluation.
Figures~\ref{fig:targeted_defenses} and~\ref{fig:untargeted_defenses} show defenses analysis for the targeted and untargeted attacks, respectively.
The `no defense' condition is a weighted mean of the relevant conditions presented in Figure~\ref{fig:attacks}. 
When comparing between targeted and untargeted attacks (Figure~\ref{fig:targeted_defenses} vs. Figure~\ref{fig:untargeted_defenses}), it can be observed that the dropout defense is significantly less effective for the targeted attacks, while the bit-depth squeezing and adversarial-training defenses are less sensitive to the attack type. 
The top-$1$ measure, which was useless when comparing different attacks, can now actually produce a good distinction between different defenses.

Figure~\ref{fig:untargeted_defenses} shows that the bit-depth squeezing and adversarial-training defenses recover more than 50\% of the classification decisions, while the dropout defense is only successful for about 20\% of the AEs.
When considering targeted attacks (see Figure~\ref{fig:targeted_defenses}), the gap between the above defenses is even wider, as the dropout defense recovers less than 10\% of the classification decisions. 
Contrary to that, the top-$5$ measure fails to differentiate some defenses from the `no defense' condition, and overall assigns rather similar scores to all the experimental conditions (Figure~\ref{fig:untargeted_defenses}).

Our two new suggested metrics again follow the more informative trend, and show that the bit-depth squeezing and adversarial-training defenses are stronger than the dropout defense under the current setting and parameters.
However, the bit-depth defense comes at a cost of a greater decrease in the accuracy of the NN in benign settings; the reduced information cancels much of the adversarial perturbations, but it also includes a loss of valuable information that is necessary for correct classification.
The same trade-off between benign and adversarial accuracy also applies to the dropout defense, however its analysis is more complicated due to the possibility of different combinations of dropout rates at different layers of the NN.
Adversarial-training, on the other hand, does not inherently suffer from such drawback; under the right setting, it was found to improve both the robustness and the generalization of NNs~\cite{wu2017adversarial}. 
However, its implementation requires more computational resources and meta-parameters tuning. 

Interestingly, $\mathit{DRR}_5$ scores are sometimes unusually lower than $\mathit{NDCG}_1$ scores.
This can be explained by the inclusion of the confidence in the $\mathit{DRR}$ score calculation; when the defense leads to a correct top-$1$ prediction, the $\mathit{NDCG}_1$ assigns the maximal score, while the $\mathit{DRR}_5$ may sometimes assign a reduced score depending on the decrease of the NN's confidence in its prediction.

\section{Conclusion}
\label{sec:conc}

As research constantly improves defense and attack methods, the need to adequately compare and evaluate them arises. 
In this paper we introduce two novel metrics, NDCG and DRR, for the comparison of attacks/defenses for multi-class pattern recognition.

Our suggested metrics are particularly valuable in cases where there are many categories with partially overlapping features; for example, different types of dogs or vehicles.
When the categories are distinctive (without any overlapping features), all successful attacks must be strong enough to convert the image label from one category to another. 
When the categories overlap, as for example is the case with different types of shotguns, weak attacks are likely to cause misclassification with a category of the most obtainable adjacent decision boundary (e.g., a shotgun of a different type, as opposed to a specific animal or landscape).
However, stronger attacks are able to deeply disturb the output of the NN in a way that will require more effort to recover (e.g., all top-ranked predicted categories now are of animals and landscapes, while shotguns receive very low probabilities).
This distinction between attacks in term of their strength may have significant implications, especially when evaluating defenses against attacks.
If the quality of the attacks is neglected, the ensemble of attacks on which the defense evaluation is based may be biased, e.g., include only weak attacks.
For example, evaluating the defense of an airport image recognition system~\cite{gota2020threat} relying on weak attacks may give a false impression of protection; strong attacks may still cause the system to classify the shotgun as a completely different category, even though weaker attacks will not succeed to switch the classification decision from one type of shotgun to another.
Of course, labeling attacks as weak or strong is neither binary nor discrete, and the effect of an attack on the NN output may have multidimensional aspects.
However, the currently used metrics for attack evaluation are either very limited (e.g., the accuracy metric) or indirect (e.g., measuring the computational effort to create the attack).
Our suggested metrics display higher sensitivity and informativeness; for example, they can detect the effect of failed attacks. 
At the same time, they still comply with prior expectations of stronger or weaker attacks, i.e., attacks that are considered strong or weak are detected as such by the metrics. 
As a consequence, the proposed metrics enable to better quantify the complicated effect of attacks on the NN output.

In order to capture as much of the above effect as possible, we based our metrics on measurements of ranked results, which originated in the field of information retrieval.
As opposed to correlation, these metrics consider the rank of each item in the list when measuring the similarity between two lists.
This feature is important in the context of attacks and defenses, since the top-ranked results of a NN are usually much more relevant than the rest of the list, and should therefore be weighted accordingly.
Measuring the relevance of categories with natural language processing techniques, such as word2vec, is also insufficient, since these techniques only consider the general semantic similarity between categories, without the unique (and in the case of image classification, visual) context of the category in each individual input.
Therefore, our metrics are based on comparison between the original predictions-list in the benign setting and the predictions-list in case of attack/defense.
As such, they realize an oracle approach, in which the evaluation is performed while having the full knowledge of the attack or defense.
Therefore, the metrics cannot be used for risk evaluation of a NN against unknown attacks.
Rather, our goal is to enable an adequate comparison between different types of attack or defenses.

As opposed to traditional ranked predictions, the items of the NNs' prediction-list are not only ranked but also weighted with a probability value, which can be interpreted as the confidence level of the NN in the prediction.
We therefore formulated the new metrics to also account for these output values, which carry additional information about the changes in the functionality of the NN as a result of the attack/defense.
Our evaluation shows that the new metrics are able to detect both attack-attack separation and attack-benign separation, contrary to the commonly-used accuracy metrics.

We demonstrated the utility of our suggested metrics in the domain of image-recognition, but they can be used in any multi-class settings, such as audio, finance, or medical data. 
Even though the two metrics aim for the same evaluation objective, they differ in their underlying reasoning.
Therefore, they can be considered as complementary, and future work may include a unified metric which will combine their two perspectives into one general approach.

\section*{Acknowledgment}
This work was supported by the Ariel Cyber Innovation Center in conjunction with the Israel National Cyber Directorate in the Prime Minister's Office.
\ifCLASSOPTIONcaptionsoff
  \newpage
\fi

\bibliographystyle{IEEEtran}
\bibliography{IEEE_TIFS}

%








\end{document}